\documentclass[aps,prd,groupedaddress,preprint]{revtex4}
\usepackage{latexsym,amsfonts}
\usepackage{hyperref}

\usepackage[utf8]{inputenc} 
\usepackage[russian]{babel} 

\begin{document}  

\title{Метод ядра эволюционного уравнения в теории гравитации
\footnote{\href{http://www1.jinr.ru/Pepan_letters/panl_2021_1/01_Gusev.pdf}{Письма в ЭЧАЯ. 2021. Т. 18, № 1(233). С. 5–12}}\\
The method of the kernel of the evolution equation in the gravity theory
\footnote{Physics of Particles and Nuclei Letters, 2021, Vol. 18, No. 1, pp. 1-4}}
\author{Юрий Владимирович Гусев\\ 
Yuri Vladimirovich Gusev}
\affiliation{Институт гравитационной физики им. А. Эйнштейна, Общество Макса Планка, Ам-Мюленберг 1, Гольм, Потсдам D-14476, Германия}
\affiliation{Центр физических исследований им. П.Н. Лебедева, Российская Академия Наук, Ленинский пр. 53, стр. 11 (38),  Москва 119991, Россия\\
{\tt Email: yuri.v.gussev@gmail.com}}

\begin{abstract}
Метод 'ковариантной теории возмущений' позволяет вычислить нелокальное ядро эволюционного уравнения на спиновом римановом многообразии. Предложенное аксиоматическое определение эффективного действия вводит в безразмерную математическую теорию универсальный масштабный параметр с размерностью квадрата расстояния. Показано, что этот чисто геометрический результат имеет физический смысл действия теории поля, включая гравитацию.  Два низших тензорных порядка этого ковариантного функционала не зависят от вида спиновой группы и локальны; они воспроизводят действие общей теории относительности с космологической постоянной. Значение универсального масштаба расстояния может определяется измеренной постоянной Хаббла. Данный масштабный параметр, рассматриваемый как физическая переменная, позволяет строить космологическую теорию аксиоматически.

The method of 'covariant perturbation theory' allowed for the computation of the kernel of the evolution equation on a spin Riemannian manifold. The proposed axiomatic definition of the effective action introduces the universal scale parameter, with the length square dimensionality, into a dimensionless mathematical theory. It is shown that this geometrical result has a physical meaning of the action of field theory, including gravity. Two orders lowest in a tensor rank in this functional are independent of a spin group and local. They reproduce the action of relativity  with the cosmological constant. The modern value of the universal scale could be determined by the measured Hubble constant. The variable scale parameter could let us build axiomatic cosmological theories.
\end{abstract}

{\bf Ключевые слова}: теория гравитации, оператор Дирака, эволюционное уравнение, эффективное действие, постоянная Хаббла, универсальный масштаб.

{\bf Keywords}: gravity theory, Dirac operator, evolution equation, effective action, Hubble constant, universal scale.

{\bf PACS}: 02.40.Ky, 04.02.Cv, 06.20.Jr, 12.10.-g, 98.80.Jk.

\date{\today}

\maketitle

В последние два десятилетия в математике активно развивалась теория геометрических потоков \cite{Ricci-flow-1-book2007}, частным видом которых является поток Риччи \cite{Andrews-book2011}. Дифференциальное уравнение потока Риччи связывает геометрические величины риманова многообразия, метрику и тензор Риччи, но при этом не содержит ковариантных производных. Потоки Риччи были использованы Г. Перельманом при доказательстве гипотезы Пуанкаре \cite{Tao-arxiv2006}, которая рассматривает многообразия постоянной кривизны, и поэтому не требует знания градиентной формы ядра эволюционного уравнения. Эта  область геометрического анализа была заложена Х. Рузом \cite{Ruse-PLMS1931}. Последующий фундаментальный вклад Дж. Синга \cite{Synge-book1960} был активно использован в физике, в частности, при неудачных попытках построения квантовой теории гравитации \cite{DeWitt-book1964} и в успешных алгоритмах спутникового ориентирования \cite{Bahder-AJP2001}. К сожалению, эти работы оказались не востребованы в геометрии, и интерес к ним возник только недавно \cite{Lee-book2019}. 

Градиентная форма потока Риччи давно известна в физике \cite{CPT2} под именем 'следа ядра уравнения теплопроводности'. Важно понимать, что эволюционное уравнение не имеет ничего общего с уравнением теплопроводности и представляет новый вид дифференциальных уравнений. Данная работа предлагает взглянуть на функциональный след ядра эволюционного уравнения как геометрический объект, отбросив  ошибочное представление о нем, как методе квантовой теории поля, укоренившееся с 1970-х. Мы разъясним физический смысл и применение ковариантного эффективного действия, вычисленного на основе эволюционного ядра \cite{Gusev-NPB2009}.

Начнём с исторической заметки о том, что действие теории гравитации (действие Гильберта-Эйнштейна) было аксиоматически получено Д. Гильбертом \cite{Hilbert-Gottingen1915} из общих принципов теории инвариантов, созданной им самим. Сегодня это действие признано основой общей теории относительности (теории гравитации) А. Эйнштейна  \cite{Dirac-book2005}. Попытки связать этот геометрический результат с остальными разделами физики никогда не прекращались и известны под именем объединённой теории поля. Единственный непротиворечивый путь развития физической теории состоит в геометризации физики, начатой Б. Риманом, У. К. Клиффордом и А. Пуанкаре. Эволюционное уравнение ведёт нас именно по этому пути. 

В то же время, метод эффективного действия позволяет нам естественным образом разрешить проблему возникновения физического масштаба (размерности) в безразмерной (математической) физической теории. Ниже показано, что математически корректное определение главного функционала физической теории, ковариантного эффективного действия, вводит в физику {\em универсальный} масштабный параметр. Такое действие принимает геометрическую форму. Данный вывод достигается методом 'ковариантной теорией возмущений'  \cite{CPT2}, которая связана исторически с операторным анализом, созданным О. Хэвисайдом \cite{Heaviside-PRSL1892}.

В ноябре 2018 г. мировое метрологическое сообщество приняло резолюцию об изменениях  в системе физических единиц {\em СИ} \cite{Stock-Metro2019}. С мая 2019 г. фундаментальные физические постоянные имеют точные фиксированные значения, а  единицы физических величин определяются этими постоянными. Так постоянная Планка стала константой, определяющей единицу массы. С введением двух новых физических постоянных М. Планком в 1900 г. \cite{Planck-AdP1900}  их число тогда стало равно числу физических единиц, породив возможность выбора значений постоянных. Если одна из констант - гравитационная постоянная, а их значениям присвоены единицы, то такую систему называют 'планковской' \cite{Planck-book1914}. В такой системе значения физических единиц, выраженные через традиционные единицы СИ, принимают непривычные значения. Считается, что эти значения обозначают пределы, где известные законы физики перестают действовать, но это не так. Вместо 1 могут быть выбраны любые другие числа, которые породят произвольно другие 'планковские' значения.

Новая СИ построена иерархически, в ней есть семь определяющих ('фундаментальных' в отличие от 'производных') постоянных, которые, тем не менее, зависят друг от друга (определяются через другие) \cite{BIPM-NewSI}. Единственная постоянная, независящая ни от какой другой - это атомная частота, определяющая единицу времени - секунду, как величина обратная частоте, выраженной целым числом. Таким  образом, в основе всей современной физики лежат натуральные числа. 

Все физические теории обладают обобщенным оператором второго порядка, который можно привести к виду \cite{CPT2},
\begin{equation}
     \hat{F}(\nabla)= \Box \hat{1} +  \hat{P} - \frac{1}{6}R\hat{1},
     \label{operator}  
\end{equation}
где присутствие члена со скаляром кривизны Риччи $R$ обусловлено историческими причинами, а сигнатура метрики - евклидова. Оператор Лапласа-Бельтрами в (\ref{operator}) построен из ковариантных производных, $\Box \equiv g^{\mu\nu} \nabla_{\mu} \nabla_{\nu}$,  содержащих как гравитационную связность, так и связность калибровочных полей, которые явно ниже не рассматриваются, но матричные обозначения, $\hat{1}$, сохраняются. Тензор напряженности калибровочных полей определяется  коммутатором ковариантных производных. Вместе с тензором кривизны Риччи, $R_{\mu\nu}$, и потенциальным членом, $\hat{P}$ эти напряженности физических полей обозначаются как $\Re$ и называются условно 'кривизнами'.

Фундаментальное уравнение геометрического анализа называется эволюционным уравнением и имеет форму \cite{CPT2},
\begin{equation}
	\frac{\mathrm{d}}{\mathrm{d} s}
	\hat{K} (s| x,x')
	=\hat{F}(\nabla^x)
	\hat{K} (s| x,x').         \label{evoleq}
\end{equation} 
Вместе с  начальными условиями
\begin{equation}
	\hat{K} (s| x,x')
	=\hat{\delta}(x,x'), \ {\sigma(x,x')/s} \gg 1, \label{delta}
\end{equation}
уравнение (\ref{evoleq}) позволяет находить эволюционное ядро, $\hat{K} (s| x,x')$,
где $\sigma(x,x')$ - мировая функция Руз-Синджа \cite{Synge-book1960}. Как показано ниже, эволюционное  ядро $\hat{K} (s| x,x')$  порождает действие изучаемой теории поля. Фундаментальное решение для эволюционного ядра задаётся ковариантной дельта-функцией (\ref{delta}). Параметр собственного времени, $s$, с физической размерностью квадрата расстояния является дополнительной переменной физической теории \cite{Fock-pt1937}, рассматриваемой в пространстве-времени с переменными $x^{\mu}$ и с размерностью $D$. Поскольку производная  первого порядка берётся по собственному времени, эволюционное уравнение позволяет получать физическое действие в ковариантной форме. В данной работе предлагается рассматривать уравнение (\ref{evoleq}) как фундаментальное уравнение теоретической физики, построенной геометрическими методами.

Ковариантное эффективное действие теории поля, включая гравитационную теорию, задаётся функциональным следом эволюционного ядра, $ {\textrm{Tr}} K (s)  =
\int {\mathrm d}^{D}  x \,  {\textrm{tr}}\,  \hat{K} (s|x,x)$, где $\textrm{tr}$ обозначает матричный след по внутренним степеням свободы, и  выполнено интегрирование по пространству-времени $\mathbb{R}^D$. В отличие от эволюционного ядра, функциональный след ${\textrm{Tr}} K (s)$ - безразмерный функционал. Ковариантная теория возмущений \cite{CPT2} даёт эволюционное ядро в асимптотически плоском пространстве-времени  виде суммы нелокальных тензорных инвариантов. При этом первые два члена этой суммы локальны (как показано прямым вычислением \cite{CPT2} и очевидно из размерных соображений),
\begin{equation}
\mathrm{Tr}\, K(s)=
\frac{1}{s^{D/2}} \int \! {\mathrm d}^{D} x\, g^{1/2}(x)
{\rm tr} \left\{ \hat{1} + s \hat{P}
	+ {\rm O}[\Re^2] \right\}  \label{TrK3}
\end{equation}
а начиная со второго порядка слагаемые этой суммы нелокальны \cite{CPT2} и в данной работе не рассматриваются. Вычисления начинаются с формального разбиения на оператора (\ref{operator}) на два нековариантных слагаемых, но перевод найденного решения для  ${\textrm{Tr}} K (s)$ в ковариантную форму выполняется с помощью нелокальных непертурбативных подстановок \cite{CPT2}. Поэтому ковариантное выражение для $\textrm{Tr}\, K(s)$ есть не ряд теории теории возмущений, а сумма нелокальных тензорных инвариантов \cite{BGVZ-JMP1994-bas}. Полученные решения \cite{CPT2,Gusev-NPB2009} справедливы при размерности пространства-времени $D<6$, но ковариантное эффективное действие ниже вычислено в четырех измерениях, соответствующих наблюдаемому физическому миру. 

Мы определим эффективное действие {\em аксиоматически},
\begin{equation}
-W (l^2) \equiv
 \int_{l^2}^{\infty}\! \frac{{\mathrm d} s}{s}\,  
	\mathrm{Tr} K(s),        \label{covaction}
\end{equation}
и будем считать, что функционал $W$ задан с точностью до произвольного множителя, значение которого находится из эксперимента. Очевидно, что у интеграла по собственному времени (\ref{covaction}) обязан быть  нижний предел, принимающий {\em произвольное} положительное значение, поскольку подынтегральный функционал не существует при $s=0$. После подстановки решения (\ref{TrK3}) в определение (\ref{covaction}) и интегрирования по $s$ получаем {\em безразмерный} функционал $W(l^2)$, явно зависящий от значения собственного времени на нижнем пределе, $l^2$,
\begin{equation}
-W(l^2)=
\sum_{n=0}^{\infty}
(l^2)^{(n-2)} 
W_{(n)}(l^2).  \label{Wmu2}
\end{equation}

Параметр  $l^2$ имеет реальный смысл в терминах физических наблюдаемых. Ковариантное эффективное действие (\ref{Wmu2}) вычислено в \cite{CPT4}, но здесь нас интересуют два его простейших члена, которые были упущены в \cite{CPT2,CPT4},
\begin{equation}
-W(l^2) =\int\! \mathrm{d}x^4\, g^{1/2}(x) \, \mathrm{tr}\,
		\Big\{
	l^{-4} \, \frac{1}{2} \hat{1} + l^{-2}\, \hat{P}
	+ {\rm O}[\Re^2] \Big\}.  \label{covaction3}
\end{equation}
Хотя эффективное действие вычислено в евклидовом пространстве-времени, его локальные члены (\ref{covaction3}) не зависят от сигнатуры метрики.

Первый член в (\ref{covaction3}) универсален для любой теории с оператором вида (\ref{operator}), а второй задается конкретной формой $\hat{P}$. В  современной физике фундаментальные поля описываются безмассовыми спинорами \cite{PDG-PRD2018}, в геометрическом описании их можно рассматривать как свойства спинового многообразия \cite{Penrose-book1984,Friedrich-book2000}. Ковариантный оператор Дирака в форме (\ref{operator}) содержит скаляр кривизны Риччи с коэффициентом $(-1/4)$ \cite{Schroedinger-GRG2020,DeWitt-book1964,Friedrich-book2000}. Тогда, для того чтобы получить эффективное действие такой теории, в  результате общего вида (\ref{covaction3}) нужно сделать подстановку, $\mathrm{tr} \hat{P} =  - \frac{1}{12} R \, \mathrm{tr}  \hat{1}$ (в которой операция матричного следа обращает тензор калибровочных полей в ноль и делает зависимость от спиновой группы тривиальной, $\mathrm{tr} \hat{1}$). Действие (\ref{covaction3}) можно привести к форме, принятой в общей теории относительности \cite{Dirac-book2005}, умножением на $12 l^2$, что согласно основной гипотезе данной работы не должно изменять физическое содержание действия (если не рассматривать космологические теории),
\begin{equation}
	\bar{W} (l^2) =   \int\! \mathrm{d}x^4\, g^{1/2}
 	\Big\{\mathrm{tr}\hat{1} \, ( 6 l^{-2}   - R)  + l^2 {\rm O}[\Re^2]
	 \Big\}.  \label{gravity}
\end{equation}
Первый член выражения (\ref{gravity}) очевидным образом интерпретируется как 'космологическая постоянная'. Подчеркнём, что это условное название, так как космологическую теорию мы не строим, а масштабный параметр будет входить во {\em все} уравнения физической теории, потому что именно $l^2$ задаёт физические размерности по иерархическому принципу, применённому в Новой СИ (2019) физических единиц \cite{BIPM-NewSI}. Второй член (\ref{gravity})  имеет форму гравитационного действия Гильберта-Эйнштейна с правильным знаком.

Можно найти значение универсального масштабного параметра, зная космологическую постоянную,
\begin{equation}
\Lambda= 6/l^2.
\end{equation}
'Стандартная космологическая модель' \cite{PDG-PRD2018} предполагает значение $\Lambda \approx 10^{-52}\, \textrm{м}^{-2}$, но, поскольку $\Lambda$ определяется постоянной Хаббла, $H_0 =  73.48 \pm 1.66\ (\textrm{км}/\textrm{с})/\textrm{Мпк}$ \cite{Riess-AJ2018}, физическая размерность которой - {\em частота}, 
\begin{equation}
H_0 \approx 2.38 \pm 0.05 \cdot 10^{-18}\ \textrm{с}^{-1}, 
\end{equation}
то естественно использовать значение радиуса Хаббла,
\begin{equation}
 l \approx c/H_0 \approx 1.26 \cdot 10^{26}\ \textrm{м}, 
\end{equation}
которое задаётся наблюдаемой $H_0$. Как $l$, так и $\Lambda^{-1/2}$  по порядку величины равняются размеру {\em наблюдаемой} Вселенной, как впервые предположил П.А.М. Дирак \cite{Dirac-book2005}. Задание универсального масштаба самым большим расстоянием в Природе при отсутствии самого маленького делает теорию  {\em наблюдаемых} физических явлений замкнутой.  

Поскольку собственное время является  параметром с физической размерностью, то масштаб $l^2$, очевидно, может считаться физической константой, только когда не принимается во внимание эволюция физического мира как целого (Вселенной). Действительно, в существующих космологических теориях постоянная Хаббла, так же как и определяемая ею космологическая постоянная рассматриваются как переменные величины  \cite{PDG-PRD2018}. Переменность универсального масштабного параметра, который иерархически задает {\em все} остальные определяющие физические константы \cite{Stock-Metro2019}, делает их тоже {\em переменными}. Это с необходимостью означает, что гипотеза Дирака об изменении гравитационной постоянной Ньютона верна \cite{Dirac-book1978}. 

Если мы принимаем три положения, которые убедительно следуют из множества известных математических и физических фактов: 1) эволюционное уравнение есть фундаментальное уравнение физики, 2) современная метрологическая система физических единиц SI (2019) самосогласованно описывает  структуру наблюдаемых физического мира, 3) космология, как наука об изучении эволюции физического мира, может быть построена на физических экспериментах, проведенных только в локальной части Вселенной, то все физические константы обязаны изменяться вместе с эволюцией Вселенной. Третий постулат, однако, не может быть экспериментально подтвержден, а потому любая космологическая теория является только научной гипотезой.

В предложенной физической теории значение космологической постоянной нельзя вычислить, эта величина может быть только измерена. Действие теории гравитации содержит как хорошо известные члены низших порядков \ref{gravity}, так и более высокие порядки в форме нелокальных тензорных инвариантов \cite{Mirzabekian-PLB1996}, которые являются членами, модифицирующими общую теорию относительности.  

После окончания представленного анализа (2016) мы нашли в литературе, что идея построения физической теории с переменным параметром типа космологической постоянной была предложена ещё П.А.М. Дираком \cite{Dirac-PRSA1973}. Дирак модифицировал теорию Г. Вейля и показал, что электромагнитное действие и космологическая постоянная возникают в теории поля из требования инвариантности действия по отношению к расширенному классу преобразований пространства-времени. Выше мы основывались на математическом принципе ядра эволюционного уравнения (\ref{evoleq}) как фундаментального уравнения физики, который ведет к более общей физической теории. Одна из задач, которые могут быть решены этим методом, состоит в аксиоматическом нахождении действия гравитационной теории \cite{Gusev-NPB2009} с целью его экспериментальной проверки. 

Более важным следствием, однако, является окончательное построение ковариантной электродинамики с масштабным параметром типа космологической постоянной. Аксиоматически определенное ковариантное эффективное действие (\ref{covaction3})  является  функционалом физических полей.  Оно находится исключительно средствами геометрического анализа и не имеет отношения к квантовой теории поля. Поскольку эффективное действие выражается через тензоры наблюдаемых полей, при варьировании по метрике оно порождает нелокальный тензор энергии-импульса \cite{MirzVilk-AP1998}, который позволяет решать уравнения эволюционных задач с начальными условиями, в частности, задачи об излучении. Напомним, что первоначально данный метод предназначался для решения проблемы Швингера о порождении частиц электромагнитным полем и проблемы об излучении Хокинга в физике чёрных дыр. Однако, математика универсальна, и поэтому ядро эволюционного уравнения применимо и в космологии, и в физике конденсированного состояния материи \cite{Gusev-FTSH-RJMP2016}. Множество задач ещё ожидают своего решения.


\end{document}